\documentclass{ws-ijmpd}
\begin{document}

\markboth{Thibault Damour, Orchidea Maria Lecian}
{About the Statistical Properties of Cosmological Billiards}

%
\catchline{}{}{}{}{}
%

\title{ABOUT THE STATISTICAL PROPERTIES\\ OF COSMOLOGICAL BILLIARDS}

\author{THIBAULT DAMOUR}

\address{Institut des Hautes Etudes Scientifiques, 35, route de Chartres, 91440 Bures-sur-Yvette, France, and\\
ICRANet, Pescara, Italy\\
damour@ihes.fr}

\author{ORCHIDEA MARIA LECIAN}

\address{Institut des Hautes Etudes Scientifiques, 35, route de Chartres, 91440 Bures-sur-Yvette, France, and\\
ICRANet, Pescara, Italy\\
lecian@ihes.fr}

\maketitle

\begin{history}
\received{Day Month Year}
\revised{Day Month Year}
\comby{Managing Editor}
\end{history}

\begin{abstract}
We summarize some recent progress in the understanding of the statistical properties of cosmological billiards.
\end{abstract}


\section{Introduction}	
A remarkable achievement of theoretical cosmology has been the construction, by Belinski, Khalatnikov and Lifshitz (BKL), of a general solution to the $4$-dimensional vacuum Einstein equations in the vicinity of a spacelike (``cosmological'') singularity \cite{KB1969a}, \cite{Khalatnikov:1969eg} \cite{BK1970}, \cite{BLK1971}. They found that this solution exhibits a never-ending oscillatory behavior, with strong chaotic properties. They could describe in detail the statistical properties of this never-ending oscillatory behavior by approximating the Einstein field equations (near the singularity) by a system of ODE's for three variables $a$, $b$, $c$ (`anisotropic scale factors'), namely
\begin{subequations}\label{abc}
\begin{align}
&2\frac{d^2\ln a}{d\tau^2}=(b^2-c^2)^2-a^4,\\
&2\frac{d^2\ln b}{d\tau^2}=(c^2-a^2)^2-b^4,\\
&2\frac{d^2\ln c}{d\tau^2}=(a^2-b^2)^2-c^4,
\end{align}
\end{subequations}
where $d\tau=-dt/(abc)$, and by approximately reducing the continuous dynamics of $a$, $b$, $c$ to a sequence of discrete maps. The crucial discrete map introduced by BKL relates the `Kasner exponents' $p_a$, $p_b$, $p_c$ describing the (approximately linear) $\tau$-evolutions of the three scale factors $a$, $b$, $c$ during two successive `epochs' (i.e. two successive segments of the dynamics (\ref{abc}) during which the influence of the right-hand side is negligible). Later studies have refined the description of the statistical properties of the chaotic BKL oscillations, notably by introducing and studying more complete discrete iteration maps (involving several real variables), and notably by a two-dimensional discrete map \cite{Chernoff:1983zz}, \cite{sinai83}, \cite{sinai85}.\\
Separately from the work of BKL, and, with a different aim and motivation, Misner realized that generic Bianchi IX homogeneous cosmological models have a ``very complex singularity'' \cite{Misner:1969hg}: a reformulation of this dynamics \cite{Misner:1974qy}, \cite{chi1972} led to the simple picture of a point moving on a Lobachevsky plane and reflecting upon \textit{fixed} billiard-type cushions. This led Chitre to remark that the dynamics of the system point is ergodic and mixing, with unique invariant Liouville measure. The problem of relating the statistical properties of the discrete BKL map to the invariance of the Liouville measure, in the continuous billiard dynamics, \`{a} la Misner-Chitre has been considered in some detail by Kirillov and Montani \cite{Kirillov:1996rd}.\\
The description of cosmological singularities in term of \textit{billiards} in (higher dimensional) Lobachevsky (or Lorentzian) spaces has recently received a new impetus from the discovery that the billiard chambers corresponding to many interesting physical theories can be identified with the ``Weyl chambers'' of certain (infinite-dimensional) Lorentzian Kac-Moody algebras \cite{Damour:2000hv}, \cite{Damour:2001sa}, \cite{Damour:2002fz}. This has raised the conjecture that, hidden below the BKL ``chaos'', there lies a remarkable ``Kac-Moody symmetry'', akin to the duality symmetries of supergravity and string theories \cite{Damour:2002cu}, \cite{Damour:2002et}, \cite{hps2009}. More precisely, in Lorentzian space, the Lagrangian $\mathcal{L}$ for these models reads
\begin{equation}
\mathcal{L}=\tfrac{1}{2}G_{ab}\dot{\beta}^a\dot{\beta}^b-V(\beta)
\end{equation}
where $G_{ab}d\beta^ad\beta^b=\sum_{a}\left(d\beta^a\right)^2-\left(\sum_ad\beta^a\right)^2$, and where the potential term $V(\beta)=\sum c_A e^{-2w_A(\beta)}$ determines the billiard walls $w_A(\beta)$. In the case of pure gravity, the only walls that can appear are either the gravitational walls $w^g_{abc}(\beta)=\beta^a-\beta^b-\beta^c+\sum\beta$ or the symmetry walls $w^{sym}_{ab}(\beta)=\beta^b-\beta^a$. In $4=3+1$ dimensions, the usual BKL billiard table is obtained by considering the three gravitational walls $w^g_{123}=2\beta^1$, $\alpha_{231}=2\beta^2$, $\alpha_{312}=2\beta^3$. This 'big billiard' corresponds to considering a \textit{diagonal metric}, described by the three scale factors $a=e^{-\beta^1}$, $b=e^{-\beta^2}$, $c=e^{-\beta^3}$. The more general case of a \textit{non-diagonal} metric corresponds to a 'small billiard' enclosed by the gravitational wall $\alpha_{123}=2\beta^1$, together with the two leading symmetry walls $w^{sym}_{12}=\beta^2-\beta^1$ and $w^{sym}_{23}=\beta^3-\beta^2$. Projecting the $3$-dimensional Lorentzian $(\beta)$ space on a Lobachevsky plane allows one to obtain some useful representations of the billiard dynamics. \\
We focus here on a few results from Ref.~\refcite{dl2010}.
 
\section{Descriptions of billiards}
The dynamics on a billiard table consists of free-flight evolution and of bounces on the boundaries of the billiard. The trajectories describing the free-flight evolution are called (Kasner) epochs, while a set of trajectories taking place in the same corner of the billiard table is called a (Kasner) era\footnote{consistently with the definition given in Eq. (5.4) of Ref.~\refcite{BLK1971}.}. Six kinds of eras can be defined, according to the corner where an era takes place, and the orientation of the first epoch of the era. All the information necessary to describe the dynamics can be encoded in maps that take into account collisions on given walls.\\
\begin{figure}
\begin{minipage}{.470\textwidth}
\centering
\includegraphics[angle=0, width=.75\textwidth]{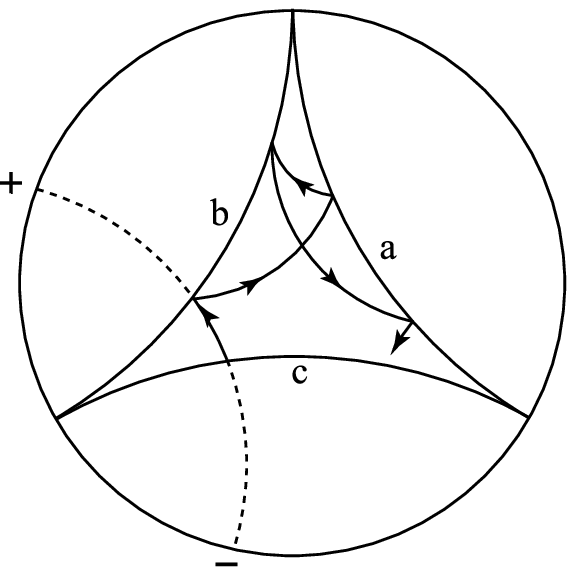}
\caption{The billiard table in the unit-disc representation. \label{fB}}
\end{minipage}
\begin{minipage}{.47\textwidth}
\centering
\includegraphics[height=1.6in]{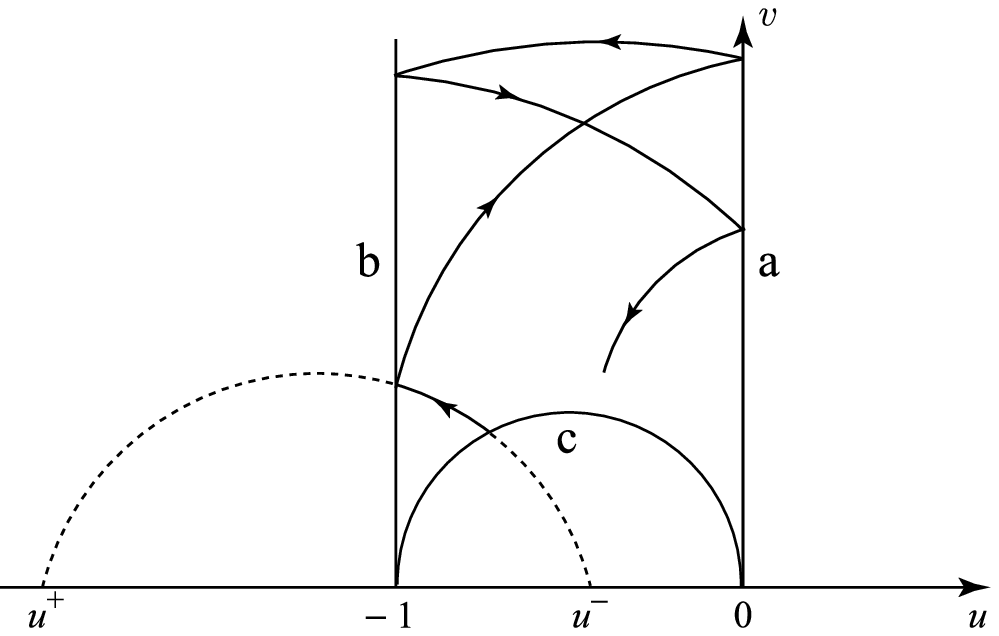}
\caption{The billiard table in the Poincar\'e half-plane representation. \label{fC}}
\end{minipage}
\end{figure}

There are two complementary descriptions of billiards: the unit disk and the Poincar\'e half plane, sketched in Fig.~\ref{fB} and Fig.~\ref{fC}, respectively.\\
The unit disk is the $2$-dimensional version of the general $n$-dimensional `ball' conformal representation, geometrically realized [within the $(n+1)$-dimensional Lorentzian $\beta$-space] by stereographically projecting (from the `South Pole' $\gamma_S$, i.e. a center of projection located on the \textit{past} unit hyperboloid) the \textit{future} unit hyperboloid onto a ($n$-dimensional) hyperplane passing through the origin in $\beta$-space. This representation is powerful in outlining the symmetry group of order six that characterizes the $3+1$-dimensional version of this model. \\
The Poincar\'e model can be obtained by a suitable geometric inversion of the ball model. The Poincar\'e half plane offers a less symmetric representation, but it offers a convenient parametrization both of the dynamics and of the oriented endpoints of the geodesics. Each collision can be described by a linear fractional transformation of the variables (of the plane):
\begin{equation}\label{3}
z'=-\frac{a\bar{z}+b}{c\bar{z}+d},\ \ ad-bc=1.
\end{equation}
For each given wall, there exists an application of the form (\ref{3}) which transforms each geodesic incident on the considered wall in the reflected geodesic. When parametrizing a geodesic by the two real numbers ($u_-, u_+$) parametrizing its end points on the 'absolute' (see Fig.~\ref{fC}), the transformation (\ref{3}) also applies, separately, to $u_-$ and $u_+$. Note, in particular, that the single real parameter $u_+$ parametrizes the Kasner exponents of the corresponding Kasner epoch. This $u_+$ parametrization of the Kasner exponents actually coincides (as emphasized by Kirillov and Montani) with the usual BKL parametrization, namely
\begin{subequations}
\begin{align}
&p_1^{BKL}(u)=-\frac{u_+}{u_+^2+u_++1},\\
&p_2^{BKL}(u)=\frac{u_++1}{u_+^2+u_++1},\\
&p_3^{BKL}(u)=\frac{u_+(u_++1)}{u_+^2+u_++1}.
\end{align}
\end{subequations}
\subsection{Integral invariants for billiards}
Several integral invariants can be constructed for billiards. Their role is to provide conserved quantities (that can be used in the statistical analysis) charactering the dynamics.\\
The usually considered energy-shell Liouville measure is a three-form obtained by reducing to the hypersurface $H=E$ the four-form $\omega^{(2)}_{PC}\wedge\omega^{(2)}_{PC}$, where $\omega^{(2)}_{PC}=dq^i\wedge dp_i-dt\wedge dH(q,p,t)$ is the (invariant) Poincar\'e-Cartan $2$-form.
However, one can also construct another invariant measure by directly restricting the Poincar\'e-Cartan two-form on any transverse section of the Hamiltonian flow. In the case of the Poincar\'e plane (asymptotic) Hamiltonian for billiards,
\begin{equation} 
H_{\gamma}(u,v,p_u,p_v)=\frac{1}{2}v^2\left(p_u^2+p_v^2\right)+V_\infty(\gamma(u,v)),
\end{equation}
the transversally reduced Poincar\'e-Cartan two-form reads
\begin{equation}\label{7}
\omega^{reduced}=\left[du\wedge d p_u+dv\wedge dp_v\right]_{v=const}=du\wedge d p_u=2\frac{du^+\wedge du^-}{(u^+-u^-)^2}.
\end{equation}
The two-form (\ref{7}) is then found (both by general arguments, and by explicit calculation) to be invariant under the discrete 'billiard map', which connects one collision to the next one. Each billiard map is given by Eq.~(\ref{3}) and is therefore of the form $u_\pm'=-\frac{au_\pm+b}{cu_\pm+d}$.\\
The invariant two-form (\ref{7}) is linked to the Liouville $3$-form by $\Omega^{(3)}_L=\omega^{(2)}_{reduced}\wedge ds$, where $ds$ is a phase-space parameter measuring the length along each geodesic segment.
\section{Big billiard hopscotch game}
\begin{figure}
\centerline{\psfig{file=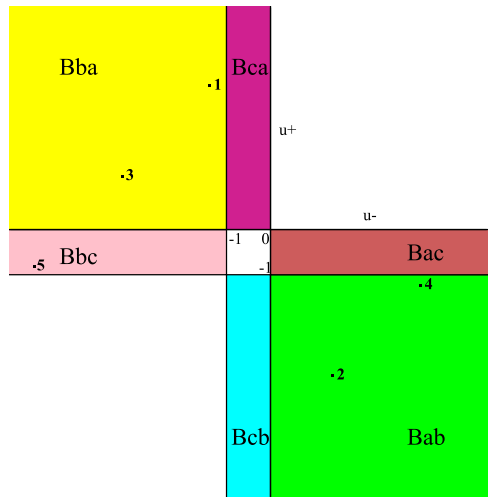,width=6.7cm}}
\vspace*{8pt}
\caption{The epoch-hopscotch court. The six different colors (shades of gray) denote the six regions of the billiard table, where the different kinds of epochs take place.  \label{fE}}
\end{figure}

We describe the dynamics of the \textit{big billiard} as a hopscotch game at different levels, such that different properties can be stressed upon.\\
The epoch-hopscotch description of the billiard dynamics is obtained by dividing the complete billiard table into six regions $B_{xy}$, each denoting the portion of the reduced phase space where all the epochs going from the wall $x$ to the wall $y$ take place. Fig.~\ref{fE} illustrates the 'jumping around' of successive epochs in the ($u^-, u^+$) plane: from the epoch number $1$ (from $b$ to $a$) to the epoch number $5$ (from $b$ to $c$). Each transformation between successive epochs is given by linear fractional transformations of ($u^-, u^+$).\\ 

\begin{figure}
\centerline{\psfig{file=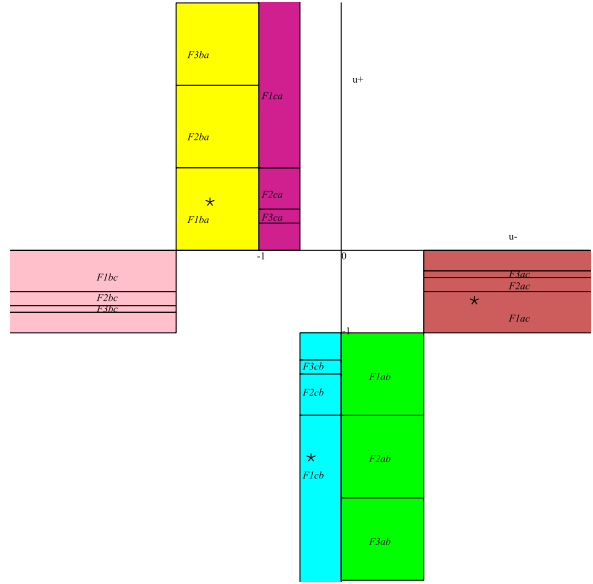,width=8.7cm}}
\vspace*{8pt}
\caption{The era-hopscotch court. The six different colors (shades of gray) denote the six regions of the billiard table, where the first epoch of the different kind of eras can take place. The asterisks indicate the one-epoch eras that form the simplest periodic configuration in the big billiard. \label{fF}}
\end{figure}

We can also use the symmetry group $S_3$ (of order $6$) to \textit{quotient} the dynamics of the big billiard. One way would be to consider a \textit{kaleidoscopic} version of the big billiard dynamics in which the single ``moving ball'' of the billiard is augmented by its $5$ (generically distinct) images under $S_3$. This leads to a billiard game where $6$ (symmetry related) balls simultaneously move within the same billiard table, and (simultaneously) bounce on its bounding walls. A second way to look at the quotiented dynamics is to replace the latter kaleidoscopic phase-space point\\ $\{q^i_{(1)},q^i_{(2)}, ..., q^i_{(6)}; p_i^{(1)}, p_i^{(2)}, ..., p_i^{(6)}\}$ by its unique representative, say $q^i_{\rm rep}$, within a fundamental domain of $S_3$, together with its corresponding momenta $p_{i \rm rep}$. When passing from the continuous billiard dynamics to the discrete billiard map from an epoch to the next epoch, the quotienting of the big billiard leads to a quotiented version of the epoch hopscotch game of Fig.~\ref{fE}. In the alternative, fundamental-domain, version of the quotiented dynamics  we could replace each $S_3$ orbit in the ($u^-,u^+$) plane by its unique representative located within, say, the box $B_{ba}$. This is achieved by means of a suitable Kasner transformation. [Indeed, the six boxes $B_{xy}$ of Fig.~\ref{fE} are exchanged under $S_3$.] In that view, the discrete quotiented big billiard would become a map from the box $B_{ba}$ onto itself. If we ignore the $u^-$ coordinate, we see that the law giving the successive values of the $u^+$ coordinate coincides with the law found long ago by BKL. This shows that the BKL discrete dynamics of the variable $u$ is obtained by: (i) quotienting our more complete hopscotch dynamics by the permutation group $S_3$, and (ii) ignoring the $u^-$ coordinate and identifying the BKL variable $u$ with $u^+$.\\
The discrete epoch hopscotch map (both in its unquotiented and in its quotiented versions) leaves invariant the measure $\omega\equiv\omega^{reduced}$, Eq.~(\ref{7}). In addition, by marginalizing $\omega$ over $u^-$, this leads to the following invariant one-dimensional measure for the usual BKL map on the $u$ parameter (in the $u>0$ convention):
\begin{equation}
w_{ba}(u^+)du^+=\frac{du^+}{u^++1}.
\end{equation}
As far as we know, this result has not been explicitly discussed before in the literature.\\
However, one cannot avail oneself of usual results of ergodic theory because the integral of $\omega$ on its total domain (e.g $B_{ba}$ for the quotiented map $\mathcal{T}_{ba}$) is \textit{infinite} (similarly the integral of $w_{ba}du^+$ is logarithmically divergent). In order to be able to use the usual results of ergodic theory, one needs a (discrete) invariant map preserving a \textit{finite} measure. This can be achieved by considering the \textit{era} hopscotch map (either in the unquotiented or quotiented forms).An \textit{era} hopscotch dynamics is obtained simply by ignoring the intermediate epochs and focusing on the map transforming the (possibly quotiented) first epoch of an era $(u^-_F, u^+_F)$ into the first epoch $(u^{- '}_F, u^{+'}_F)$ of the next era. The unquotiented era-hopscotch court is sketched in Fig.~\ref{fF} , and the explicit expression of the unquotiented era map $\mathcal{T}$ is listed in Table~\ref{t1}. In the \textit{quotiented case}, the resulting discrete (two-dimensional) map, say $\mbox{\large{$\mathsf{T}$}}$, coincides with the map studied by Chernoff-Barrow and by Lifshitz-Khalatnikov-Sinai-Khanin-Shchur. When using a representative of the $S_3$ orbit within the $B_{ba}$ box, the quotiented era map $\mbox{\large{$\mathsf{T}$}}$ is a map of $F_{ba}$ onto itself, i.e. the rectangular domain $-2<u^-_F<-1$, $0<u^+_F<+\infty$. The explicit expression of the map $\mbox{\large{$\mathsf{T}$}}$ is 
\begin{equation}\label{cblksks}
\mbox{\large{$\mathsf{T}$}}u_F^\pm=+\frac{1}{u_F^\pm-[u_F^+]}-1.
\end{equation}
This maps leaves invariant the \textit{restriction} of the two-form $\omega$ to the domain $F_{ba}$; by contrast with the original measure $\omega$ on the full hopscotch court, this restricted measure has now a \textit{finite integral}, namely $\int_{\omega_{F_{ba}}}\omega_F=2\ln2$.
\begin{table}[ph]
\tbl{The era-hopscotch transformations, which defines the big-billiard map $\mathcal{T}$.}
{\begin{tabular}{@{}cccc@{}} \toprule

   $F_{ab}$  & $n^{ab}$ odd   & $F'_{bc}$ & $u_{F_{bc}}=-u_{F_{ab}}-n^{ab}-1$  \\ 
      & $n^{ab}$ even & $F'_{ac}$ & $u_{F_{ac}}=u_{F_{ab}}+n^{ab}$ \\
  \hline
  $F_{ba}$  & $n^{ba}$ odd  & $F'_{ac}$ & $u_{F_{ac}}=-u_{F_{ba}}+n^{ba}-1$  \\
      & $n^{ba}$ even & $F'_{bc}$ & $u_{F_{bc}}=u_{F_{ba}}-n^{ba}$ \\
      \hline
  $F_{ac}$  & $n^{ac}$ odd  & $F'_{cb}$ & $u_{F_{cb}}=-\frac{1}{n^{ac}+1+\tfrac{1}{u_{F_{ac}}}}$  \\
      & $n^{ac}$ even & $F'_{ab}$ & $u_{F_{ab}}=\frac{1}{n^{ac}+\tfrac{1}{u_{F_{ac}}}}$ \\
      \hline
  $F_{ca}$  & $n^{ca}$ odd  & $F'_{ab}$ & $u_{F_{ab}}=\frac{1}{n^{ca}-1-\tfrac{1}{u_{F_{ca}}}}$  \\
      & $n^{ca}$ even & $F'_{cb}$ & $u_{F_{cb}}=-\frac{1}{n^{ca}-\tfrac{1}{u_{F_{ca}}}}$ \\
      \hline
  $F_{bc}$  & $n^{bc}$ odd  & $F'_{ca}$ & $u_{F_{ca}}=-1+\frac{1}{n^{bc}+1-\tfrac{1}{1+u_{F_{bc}}}}$  \\
      & $n^{bc}$ even & $F'_{ba}$ & $u_{F_{ba}}=-1-\frac{1}{n^{bc}-\tfrac{1}{1+u_{F_{bc}}}}$ \\
      \hline
   $F_{cb}$  & $n^{cb}$ odd  & $F'_{ba}$ & $u_{F_{ba}}=-1-\frac{1}{n^{cb}-1+\tfrac{1}{1+u_{F_{cb}}}}$  \\
      & $n^{cb}$ even & $F'_{ca}$ & $u_{F_{ca}}=-1+\frac{1}{n^{cb}+\tfrac{1}{1+u_{F_{cb}}}}$ \\
     
\botrule        
\end{tabular} \label{t1}}
\end{table}
\subsection{Recovering information about the unquotiented dynamics}
Quotienting the dynamics of the big billiard implies a loss of information with respect to the full unquotiented dynamics. We analyse here what kind of information is lost, and show how to recover this information.\\ 
Within the notation we have developed, the continued-fraction expansions\footnote{The notation $[x]$ for $x\in\mathbb{R}$ denotes the usual integer part of $x$ when $x\ge0$ (e.g. $[\pi]=3$), and $-[-x]\le0$ when $x\le0$ (so that $[-\pi]=-3$); we define the fractional part of a negative number accordingly, e.g. $\{-\pi\}=-0.14...$. Furthermore, we have denoted the first integer of the decomposition of $u^+_F$ as $n_{ba}-1=[u^+_F]$, so that $n_{ba}=[u^+_F]+1$ denotes the length of the era starting with $u^+_F$.}  of the first ($u^-,u^+$) values of a (quotiented) era starting in the box $F_{ba}$ can be written as
\begin{subequations}\label{contfrac}
\begin{align}
&u^+_F\equiv n_{ba}-1+ [ n_2,n_3,n_4,... ] \equiv [n_{ba}-1; n_2,n_3,n_4,...],\\
&u^-_F=-1-[m_1, m_2, m_3, ...].
\end{align}
\end{subequations}
 In terms of these decompositions pertaining to the first era, we can write the first $u^-,u^+$ values of the $N$-th era (with $N=2,3,...$), i.e. the ($N-1$)-th iteration of the $\mbox{\large{$\mathsf{T}$}}$ map
\begin{subequations}\label{137}
\begin{align}
&\mbox{\large{$\mathsf{T}$}}^{N-1}u^+_F=n_N-1+[n_{N+1}, n_{N+2}, n_{N+3}, ...],\\
&\mbox{\large{$\mathsf{T}$}}^{N-1}u^-_F=-1-[n_N, n_{N-1}, ..., n_2, n_{ba}, m_1, m_2, ...].
\end{align}
\end{subequations}
\begin{table}[ph]
\tbl{\label{table6} The Kasner tranformations.}
{\begin{tabular}{@{}cccc@{}} \toprule    

    $k_5$ & $u_K^{ba}=-u_{ab}-1$ \\ \hline
    $k_2$ & $u_K^{ba}=-(u_{ac}+1)/u_{ac}$ \\ \hline
    $k_3$ & $u_K^{ba}=-u_{bc}/(u_{bc}+1)$ \\ \hline
    $k_1$ & $u_K^{ba}=1/u_{ca}$ \\ \hline
    $k_4$ & $u_K^{ba}=-1/(u_{cb}+1)$ \\  
    
\botrule        
\end{tabular} }\caption{ The Kasner transformations.}
\end{table}
Furthermore, if we start an unquotiented era hopscotch dynamics in some specific region $F_{xy}$, we can first map it to the reference region $F_{ba}$ by some specific `Kasner transformation' $k_{xy}$ (listed in Table \ref{table6}) to get its $ba$-representative, $u^{+[ba]}_{F_{xy}}$, namely
\begin{equation}
u^{+[ba]}_{F_{xy}}\equiv k_{xy}u^+_{F_{xy}}\equiv n_{xy}-1+[n_2, n_3, ...].
\end{equation}
We see that the quotiented era dynamics encodes the lengths of successive eras, but ignores the fuller information contained in the original, unquotiented era hopscotch dynamics, namely the precise \textit{corner}, $\{x_N, y_N\}$ (with $x_N, y_N \in\{a, b, c\}$) and \textit{orientation}, ($x_N, y_N$), (i.e. $x_N\rightarrow y_N$), of the first epoch of the $N$-th hopscotch era, such that one could express the explicit result of iterating $(N-1)$ times the unquotiented $\mathcal{T}$ map in terms of the simpler action of the quotiented $\mbox{\large{$\mathsf{T}$}}$ map, namely 
\begin{equation}\label{uN}
u^\pm_{F_{x^{N},y^{N}}}=\mathcal{T}^{N-1}u^\pm_{F_{xy}}=\mathcal{T}\circ...\circ\mathcal{T}\circ\mathcal{T}u^\pm_{F_{xy}}=k_{x^{N},y^{N}}^{-1}\mbox{\large{$\mathsf{T}$}}^{N-1}k_{xy}u^\pm_{F_{xy}}.
\end{equation}
The missing information is recovered by considering the quantities 
\begin{subequations}\label{rhoeta}
\begin{align}
&\rho_N=e^{i\theta_N},\ \ {\rm with}\ \ \theta_N\equiv-D[k_F]\tfrac{2\pi}{3}\left(\epsilon_1+\epsilon_1\epsilon_2+...+\epsilon_1\epsilon_2...\epsilon_N\right),\\
&\eta_N=\epsilon_1\epsilon_2...\epsilon_N.
\end{align}
\end{subequations}
 where $\epsilon_j\equiv(-)^{n_j+1}$, $n_j$ denotes the number of epochs contained in the $j$-th era, and $D[k_F]$ is the determinant of the first Kasner transformation $k_{xy}$ involved in (\ref{uN}). The procedure determining $k^{-1}_{x^{N},y^{N}}$ is as follows: (i) starting from $k_{xy}$ and the continued-fraction decomposition of $k_{xy}u^+_F=n_1-1+[n_2, n_3, ...]$, one determines the $\epsilon_i$, and the $\rho_N$ and $\eta_N$; (ii) then $\rho_N$ determines the rotation (in the unit-disc representation) between $xy$ and the final corner $x_N, y_N$, and $\eta_N$ determines whether this corner is ``parallel'' ($\eta_N=1$) or ``antiparallel'' ($\eta_N=-1$) to the initial $xy$; (iii) finally, knowing the ordered corner $(x^N, y^N)$, Table~\ref{table6} determines the transformation $k_{x^N, y^N}$ that maps it onto the $ba$ corner.
\subsubsection{Periodic orbits} 
\begin{figure}
\centerline{\psfig{file=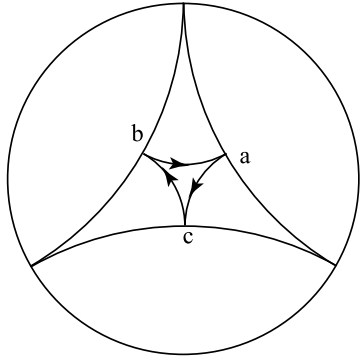,width=4.7cm}}
\vspace*{8pt}
\caption{The periodic succession of one-epoch eras of the big billiard in the unit-disc representation. \label{fA}}
\end{figure}

Periodic orbits offer another way to contrast the unquotiented dynamics with its quotiented version. First, any $m$-periodic orbit of the one-dimensional BKL map $T_{\rm BKL}$, parametrized by the special values of $u^+$ that admit a (regular) periodic continued-fraction expansion of the type
\begin{equation}
\label{eq:mcfe}
u^+ + 1 = [n_1 ; n_2 , n_3 , \ldots n_m , n_1 , n_2 , n_3 , \ldots] \, ,
\end{equation}
gives rise to a unique corresponding $m$-periodic orbit of the two-dimensional $\mbox{\large{$\mathsf{T}$}}$ map. Second, any $m$-periodic orbit of $T_{\rm BKL}$ and $\mbox{\large{$\mathsf{T}$}}$ can be lifted to some periodic orbit of period $mp$ of the full, unquotiented billiard, according to the condition
\begin{equation}
{\mathcal T}^{mp} (u^- , u^+)=k_*^p (u^- , u^+)= (u^- , u^+) \, ,
\end{equation}
where $k_*$ is a suitable Kasner transformation (which depends on $m$ and on the considered periodic orbit of $\mbox{\large{$\mathsf{T}$}}$), and $p$ is the order of $k_*$. The \textit{order} of a particular group element, such as $k_*$, is the smallest integer $p$ such that $k_*^p=k_0$; as a transposition is of order $2$, and a cyclic permutation, ($123$) or ($321$), of order $3$, we see that the order $p$ of $k_*$ must be equal to $p=1, 2$ or $3$. In other words, the period in the unquotiented billiard can be up to three times larger than the one in the quotiented one. For example, the simplest $1$-periodic orbit of $\mbox{\large{$\mathsf{T}$}}$ given by the golden ratio $u^++1=[1;1,1,1,...]=(1+\sqrt{5})/2$ corresponds to a $3$-periodic orbit of $\mathcal{T}$, as shown in Fig. \ref{fA}.
\vspace{0.1cm}



\end{document}